\documentclass[aps,showpacs,preprint]{revtex4}
\usepackage{hyperref}
\usepackage{graphicx}
\usepackage{amsmath}
\usepackage{color}

\newcommand{\be}{\begin{equation}}
\newcommand{\ee}{  \end{equation}}
\newcommand{\ba}{\begin{eqnarray}}
\newcommand{\ea}{  \end{eqnarray}}
\newcommand{\ket}[1]{\left|#1\right>}
\newcommand{\bra}[1]{\left< #1 \right|}

\begin{document}
\title{Direct observation of quantum criticality in Ising spin chains}
\author{
Jingfu Zhang\footnote{Corresponding author: j87zhang@iqc.ca;
zhangjfu2000@yahoo.com} } \affiliation{Institute for Quantum
Computing and Department of Physics, University of Waterloo,
Waterloo, Ontario, Canada N2L 3G1 }
\author{Fernando M. Cucchietti\footnote{Corresponding author: fernando@cucchietti.com}}
\affiliation{T4, Theoretical Division, MS B213, Los Alamos
National Laboratory, Los Alamos, New Mexico, 87545 USA}

\author{
C. M. Chandrashekar} \affiliation{Institute for Quantum Computing
and Department of Physics, University of Waterloo, Waterloo,
Ontario, Canada N2L 3G1 }

\author{
Martin Laforest} \affiliation{Institute for Quantum Computing and
Department of Physics, University of Waterloo, Waterloo, Ontario,
Canada N2L 3G1 }

\author{Colm A. Ryan} \affiliation{Institute for Quantum Computing and Department of Physics,
University of Waterloo, Waterloo, Ontario, Canada N2L 3G1 }

\author{Michael Ditty} \affiliation{Institute for Quantum Computing and Department of Physics,
 University of Waterloo, Waterloo, Ontario, Canada N2L 3G1 }

\author{Adam Hubbard} \affiliation{Institute for Quantum Computing and Department of Physics,
University of Waterloo, Waterloo, Ontario, Canada N2L 3G1 }

% gamble - fixed affiliation
\author{
John K. Gamble} \affiliation{Department of Physics, The College of
Wooster, Wooster, Ohio, 44691 USA}
\author{
Raymond Laflamme\footnote{Corresponding author: laflamme@iqc.ca}}
\affiliation{Institute for Quantum Computing and Department of
Physics, University of Waterloo, Waterloo, Ontario, Canada N2L 3G1
} \affiliation{Perimeter Institute for Theoretical Physics,
Waterloo, Ontario, N2J 2W9, Canada}
\date{\today}

\begin{abstract}
We use NMR quantum simulators to study antiferromagnetic Ising
spin chains undergoing  quantum phase transitions. Taking
advantage of the sensitivity of the systems near criticality, we
detect the critical points of the transitions using a direct
measurement of the Loschmidt echo. We test our simulators for spin
chains of even and odd numbers of spins, and compare the
experimental results to theoretical predictions.
\end{abstract}
\maketitle

\section{Introduction}

Quantum phase transitions (QPTs) describe sudden changes of the
ground state of a many-body quantum system as a
non-thermal control parameter moves through some
critical value \cite{QPTbook} (at zero temperature).
QPTs are relevant not only for understanding
of quantum many-body systems, but also for other problems such as quantum
entanglement \cite{entangedQPT} and quantum computing, e.g.,
adiabatic quantum computing \cite{adiabatic} and quantum
estimation \cite{Zanardi07}. Interesting phenomena related to QPTs
have recently been experimentally observed in various systems,
such as heavy fermions and Bose-Einstein condensates
\cite{RevNaturePhys08}.

There has been a recent flurry of activity following the observation \cite{sunPRL06} that
the proximity to a quantum critical point
enhances the sensitivity of a system to external perturbations,
as measured by quantum-information-theoretical quantities such
as the Loschmidt echo \cite{sunPRL06} or the ground state fidelity
 \cite{zanardi:031123}. Exploiting such
sensitivity, one can detect quantum criticality by coupling an
additional spin as a probe to the system undergoing a QPT. This was
suggested in \cite{Paz} and demonstrated in
\cite{Zhang08}, where the local coupling to the probe qubit
was used as the perturbation.

Here, we implement an alternative method to detect the critical
point of a QPT by measuring an arbitrary qubit of the quantum
critical system while applying a global perturbation. The critical
parameters of a general QPT, i.e., including critical points and
exponents, can in principle be detected using our method. Our
approach does not require an additional probe spin, which makes
the experimental implementation easier. In contrast to our method,
in the previous approach \cite{Zhang08}  the efficiency of
detection depended on the nature of the phases on both sides of
the critical points, and could be affected, or even rendered
insensitive, by the locality of the probe. Furthermore, because
our method uses a global perturbation, it increases the echo signal --
making it, in principle, better suited for scalability with the size of the system.

The paper is organized as follows: In section II we introduce the
model and discuss how we use it to simulate a second order QPT. In
section III we review the behavior of the Loschmidt echo in a
critical system using a perturbative treatment. In particular, we
discuss  the echo decay rate and its scaling near the critical
point. In section IV we describe the experimental implementation
for even and odd spin chains using nuclear magnetic resonance, and
compare our results to theoretical expectations. We offer
concluding remarks in section V.

\section{Ising chain with a tilted field}

To demonstrate the detection of quantum criticality, we simulate
the QPTs using a one-dimensional antiferromagnetic Ising model
with the Hamiltonian
\begin{equation}\label{hamIsing}
    H=\sum_{i=1}^{N-1}\sigma^{i}_{z}\sigma^{i+1}_{z}
    +B_{z}\sum_{i=1}^{N}\sigma^{i}_{z}+B_{x}\sum_{i=1}^{N}\sigma^{i}_{x},
\end{equation}
where $B_{z}$ and  $B_{x}$ denote longitudinal and transverse
magnetic fields, respectively, $\sigma_z^i$ and $\sigma_x^i$ are
Pauli matrices acting on spin $i$ of the chain, and the coupling
strength has been set to unity. This type of model has been
extensively studied in the literature in the contexts of
statistical physics \cite{IsingStat84}, quantum computing
\cite{compile}, quantum chaos \cite{qchaos}, and QPTs
\cite{pfeuty,physRepIsing,IsingQPT,Bose01Ising,Zoller05,Werner05,Peng05,Mostame07,Gu07,Ovchinnikov}.

\begin{figure}[bt]
\includegraphics[width=4in]{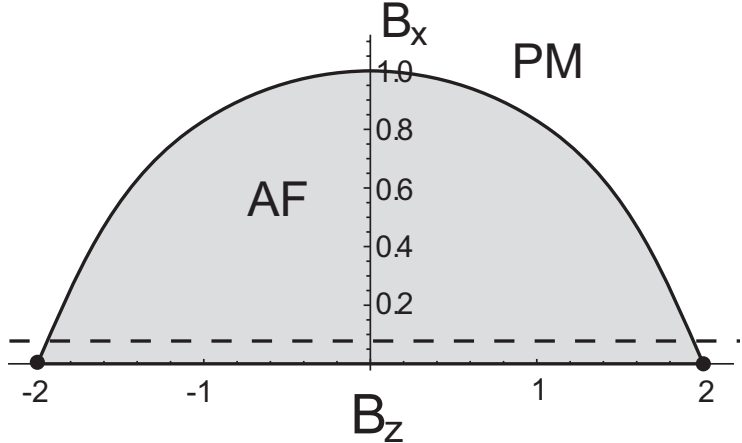}
%Matlab :phase.m and phase.fig in F:\UW\PhaseTrans
\caption{Phase diagram of the antiferromagnetic Ising chain with
transverse and longitudinal fields, $B_x$ and $B_z$ respectively,
in the thermodynamic limit of infinite chain size
\cite{IsingAF,Ovchinnikov}. The coupling strength is chosen as the
unit for $B_x$ and $B_z$, [see Eq. (\ref{hamIsing}) in text]. In
the shadowed region inside the circle the ground state is (doubly
degenerate) antiferromagnetic (AF), and in the clear region
outside it the ground state is paramagnetic (PM). The transition
line between both phases is a second order critical line, while
the points at $B_x=0$ are first order transitions. The phase
diagram corresponds to that of a two dimensional classic Ising
model with field equal to $B_z$ and effective temperature
proportional to $B_x$. The dashed line shows qualitatively the
region we explore experimentally, where the critical points in the
thermodynamical limit are close to $B_z=\pm2$. For the finite
systems used in our experiments, we need to consider boundary
effects which show up like extra sub-phases inside the AF phase.
For odd $N$, a new critical point appears at $B_z=0$, while for
even $N$ two extra critical points appear at $B_z=\pm 1$.
}\label{phasediagram}
\end{figure}

Notice that the general case of Eq. (\ref{hamIsing}) with $B_z\neq
0$ and $B_{x}\neq 0$ cannot be solved exactly using Jordan-Wigner
transformation methods because the longitudinal field maps into
high order coupling of the resulting fermions. This can also be
seen by noting that the Hamiltonian (\ref{hamIsing}) can be mapped
into a classical 2D Ising model \cite{suzuki}, with $B_z$ the
longitudinal field and $B_x$ an effective temperature -- which
means that our quantum simulation can also be seen as a simulation
of this archetypal model of classical phase transitions. The map
between a quantum $d$ dimensional spin system into a $d+1$
classical Ising system \cite{suzuki} lets us obtain the phase
diagram of Hamiltonian (\ref{hamIsing}) in the thermodynamic
limit, which corresponds to that of the 2D classical
antiferromagnetic Ising model \cite{IsingAF}, and is shown
qualitatively in Figure \ref{phasediagram}. The critical line is
second order except for $B_{x}=0$, where it is a first order
transition. As we will use only a few qubits, we are concerned
here only with finite systems. Furthermore, the Loschmidt echo
decay rate typically increases with system size \cite{Paz}, which
implies that in the thermodynamic limit the echo would decay
infinitely fast (unless the perturbation is simultaneously reduced
to zero, where a singular decay rate would be obtained
\cite{Paz}). In the finite size systems under consideration, the
gap across a second order transition never closes, but rather
reaches a minimum near the critical point (this minimum goes to
zero in the thermodynamic limit). Furthermore, for finite systems
we need to consider odd-even effects, which in our model system
will introduce "quasi"-phases that come from boundary effects and
merge in the thermodynamic limit.

Let us consider first the ground states for $B_x=0$, which will be
relevant for our experiments. We keep in mind that in this particular case
the system undergoes crossovers as a function of $B_z$, since only
the energies,  not the eigenstates, depend on $B_z$. When $N$ is
an odd integer, the ground state of the system is
%Gamble - modernized the following equation to use cases
\begin{equation}\label{odd}
    \ket{ \psi^{o}(B_z)}=\begin{cases}
    |\underbrace{00..0}_{N}\rangle & (B_z<-2)\\
    |\underbrace{01...01}_{(N-1)/2 \hspace{0.1cm} pairs
    \hspace{0.1cm} of \hspace{0.1cm} 01}0\rangle& (-2< B_z< 0)\\
    |\underbrace{10...10}_{(N-1)/2 \hspace{0.1cm} pairs
    \hspace{0.1cm}of \hspace{0.1cm} 10}1\rangle& (0< B_z< 2)\\
    |\underbrace{11...1}_{N}\rangle & (B_z> 2)
   \end{cases}
\end{equation}
where $\ket{0}$ and $\ket{1}$ are the eigenstates of $\sigma_z$.
We denote the four phases of the ground state as
$|\psi_{k}^{o}\rangle$ with $k=1$, $\ldots$, $4$. The energy of
the ground state is
\begin{equation}\label{energyodd}
    E_{g}^{o}(B_z)=\begin{cases}
    N(B_z+\frac{N-1}{N}) & (B_z\leq-2)\\
    N(\frac{B_z}{N}-\frac{N-1}{N}) & (-2\leq B_z\leq 0)\\
    N(-\frac{B_z}{N}-\frac{N-1}{N})& (0\leq B_z\leq 2)\\
    N(-B_z+\frac{N-1}{N}) & (B_z\geq 2)
   \end{cases}.
\end{equation}
We denote the energy corresponding to the four phases
$|\psi_{k}^{o}\rangle$ as $E_{g,k}^{o}$, respectively.
$B_z=B_c=\pm2$ and $0$ are the crossover points, where the system
has a degenerate ground state. $(N+1)/2$ degenerate states exist
at $B_c=\pm2$, making them the multiphase points of the system
\cite{IsingQPT,supersymmetry1}.

When $N$ is an even integer larger than $2$, the ground state of
the system is
%Gamble - modernized the following equation to use cases
\begin{widetext}
\begin{equation}\label{even}
   \ket{\psi^{e}(B_z)}=\begin{cases}
    \vspace{0.2cm}
    |\underbrace{00..0}_{N}\rangle & (B_z<-2)\\
    \vspace{0.2cm}
    \frac{1}{\sqrt{2}}\left [ |\underbrace{01...01}_{(N-2)/2 \hspace{0.1cm} pairs
    \hspace{0.1cm} of \hspace{0.1cm} 01 }00\rangle
    +|00\underbrace{10...10}_{(N-2)/2 \hspace{0.1cm} pairs
    \hspace{0.1cm} of \hspace{0.1cm} 10}\rangle \right ]& (-2< B_z< -1)\\
 \vspace{0.2cm}
    \frac{1}{\sqrt{2}}\left [|\underbrace{01...01}_{N/2 \hspace{0.1cm} pairs
    \hspace{0.1cm} of \hspace{0.1cm} 01}\rangle
    +|\underbrace{10...10}_{N/2 \hspace{0.1cm} pairs
    \hspace{0.1cm} of \hspace{0.1cm} 10}\rangle \right ]& (-1< B_z< 1)\\
 \vspace{0.2cm}
    \frac{1}{\sqrt{2}}\left [ |11\underbrace{01...01}_{(N-2)/2 \hspace{0.1cm} pairs
    \hspace{0.1cm} of \hspace{0.1cm} 01}\rangle
    +|\underbrace{10...10}_{(N-2)/2 \hspace{0.1cm} pairs
    \hspace{0.1cm} of \hspace{0.1cm} 10}11\rangle \right ]& (1< B_z< 2)\\
 \vspace{0.2cm}
   |\underbrace{11...1}_{N}\rangle & (B_z> 2)
   \end{cases},
\end{equation}
\end{widetext}
and the energy of the ground state is
\begin{widetext}
\begin{equation}\label{energyeven}
   E_g^{e}(B_z)=\begin{cases}
   N(B_z+\frac{N-1}{N}) & (B_z\leq-2)\\
   N(\frac{2B_z}{N}-\frac{N-3}{N})& (-2\leq B_z\leq -1)\\
 N(-1+\frac{1}{N})& (-1\leq B_z\leq 1)\\
 N(-\frac{2B_z}{N}-\frac{N-3}{N})& (1\leq B_z\leq 2)\\
 N(-B_z+\frac{N-1}{N}) & (B_z\geq 2)
   \end{cases}.
\end{equation}
\end{widetext}
 The crossover points are
$B_c=\pm2$ and $B_c=\pm1$. Points $B_c=\pm2$ are also multiphase
points, each with $N/2$ degenerate states. The five phases are
denoted as $|\psi_{k}^{e}\rangle$ with $k=1$, $\ldots$, $5$ and
the corresponding energy is represented as $E_{g,k}^{e}$.

From Eqs. (\ref{energyodd}) and (\ref{energyeven}), one finds that
if $N\rightarrow\infty$, $E_{g,2}^{o}\rightarrow E_{g,3}^{o}$,
$E_{g,2}^{e}\rightarrow E_{g,3}^{e}$, and $E_{g,4}^{e}\rightarrow
E_{g,3}^{e}$. Hence, in the thermodynamic limit only the
multiphase points $B_c=\pm2$ are the crossover points, and
$|\psi_{2}^{o}\rangle$, $|\psi_{3}^{o}\rangle$,
$|\psi_{2}^{e}\rangle$, $|\psi_{3}^{e}\rangle$, and
$|\psi_{4}^{e}\rangle$ are "quasi"-phases that merge into a single
antiferromagnetic phase (see Figure \ref{phasediagram}). The
finite size energy phase diagrams are shown in Figure
\ref{fphase}(a-b).

In general, second order QPTs are characterized by a closing of
the gap between the ground and first excited energy levels at the
critical points (in the thermodynamic limit). Using our small
quantum information processors, we will simulate the evolution of
the quantum system described by Hamiltonian (\ref{hamIsing}) in a
regime where its spectrum is similar to the general case of a
finite-size second order QPT (that is, with a small but finite
gap). We achieve this by using a small transverse field $B_x$ to
lift the degeneracy at points $B_c$, which makes the spectra
resemble a continuous QPT \cite{cqpt}. Thus, we explore the
transitions crossed by the dashed line in Figure
\ref{phasediagram}. In the analysis of our results we must take
into consideration finite-size effects such as the size of the gap
at the critical points, and the additional "quasi"-phases
introduced by flipping a finite number of spins at the ends of the
chain -- which makes a distinction between experiments with odd
and even chains.

\section{The Loschmidt echo and quantum phase transitions}

\subsection{Detection of critical parameters}

Let us consider a system with Hamiltonian $H_0$, controlled by an
external parameter $\lambda$ (in our experiments, $\lambda$ is the
longitudinal field $B_z$). We assume $H_0$ to have gapped phases
around a critical point $\lambda_c$, and without loss of
generality we write a perturbed system Hamiltonian
$H_1=H_0+\varepsilon V$, where $V$ is an arbitrary Hermitian
operator (to be defined later) and $\varepsilon$ is the strength
of the perturbation. Taking the ground state $\ket{0(\lambda)}$ of
$H_0$ as the initial state, the time dependent Loschmidt echo
\cite{LEReview} takes the form \be L(t)\equiv|\ell(t)|^2=|
\bra{0(\lambda)}e^{i H_1 t} e^{-i H_0 t} \ket{0(\lambda)} |^2.
\label{define_L} \ee Notice that the evolution under $H_0$ gives a
physically irrelevant phase, which we keep for convenience of
notation. The correspondence of the quantum critical points a QPT
and the minima of the Loschmidt echo for {\em long times} has been
shown for many systems \cite{sunPRL06, Rossini}. However, the
dynamical behavior for short times depends on the symmetries of
the phases around the critical point and those of the perturbation
operator. For instance, a monotonic increase of the decay rate
with a singularity in its first derivative  has been observed for
some systems with local perturbations \cite{Rossini}. On the other
hand, in the experimental results shown in the next section we
observe that, for a fixed short time, the Loschmidt echo
approaches a minima in the vicinities of the critical points. In
this section we are concerned with providing a theoretical
framework to these experimental observations. For this, we will
analize the Loschmidt echo for short times using a perturbative
approach (similar to the one of Ref. \cite{Rossini}), and
particularize to the universality of the system we simulate in the
experiments.

For small perturbations $\varepsilon$ we expand the echo amplitude
\be
\ell(t)\simeq
 \left. \ell(t)\right|_{\varepsilon=0}
 +
 \left. \frac{\partial \ell(t)}{\partial \varepsilon} \right|_{\varepsilon=0}
 \varepsilon
+
\left. \frac{\partial^2 \ell(t)}{\partial \varepsilon^2} \right|_{\varepsilon=0}
\frac{\varepsilon^2}{2}.
\label{taylor}
\ee
The first term is
\be
 \left. \ell(t)\right|_{\varepsilon=0} = \bra{0(\lambda)} e^{i H_0 t}e^{-i H_0 t} \ket{0(\lambda)} =1.
\label{mu0}
\ee
For the second and third terms, we need to
compute derivatives of the perturbed evolution operator.
We can do this by expanding into infinite
series and re-summing after computing the expectation value of the
operators in the ground state. After some algebra (see appendix \ref{appendix}), we find
\ba
 \left. \frac{\partial \ell(t)}{\partial \varepsilon} \right|_{\varepsilon=0} &  = & (-i t) V_{00} \\
\left. \frac{\partial^2 \ell(t)}{\partial \varepsilon^2} \right|_{\varepsilon=0} & =&  2
\sum_{\alpha=0}^{N-1}  |V_{0\alpha}|^2 \times \nonumber \\
& & \frac{e^{-i (E_\alpha-E_0) t}-1+i t(E_\alpha -E_0)}
{(E_\alpha-E_0)^2},
\label{moments}
\ea
where  $\alpha$ indexes the $N$ eigenstates of $H_0$ with energy $E_\alpha$,
$E_0$ is the ground state energy,
and $V_{0\alpha}=\bra{\alpha(\lambda)} V \ket{0(\lambda)}$.
The second order term of Eq. (\ref{moments}) resembles the
so called fidelity susceptibility \cite{You} and the quantum geometric tensor \cite{Geometric}
that have been shown to display singular behavior and scaling near a critical point.
Indeed, if we take the Fourier transform of $|L(t)|^2$, we obtain the fidelity susceptibility \cite{You}
for low frequencies. Higher frequency components appear that are related to the extra terms
in the local density of states that generalizes the ground state fidelity \cite{sunPRL06}.

Our final perturbative expression for the Loschmidt echo is then
\be
L(t) \simeq 1 - 2 \varepsilon^2 \sum_{\alpha=1}^{N-1} |V_{0\alpha}|^2 \frac{1-\cos (E_\alpha-E_0)t}{(E_\alpha-E_0)^2}.
\label{finalL}
\ee

\subsection{Landau-Zener QPT toy model}

When the main contribution to the sum in Eq. (\ref{finalL}) is given by the first excited state,
we can approximate
\be
L(t)\simeq 1- 2 \frac{|V_{01}|^2}{\Delta^2} \varepsilon^2 (1-\cos{\Delta t}),
\label{Lapprox}
\ee
where $\Delta=E_1-E_0$ is the gap that has a minimum at the critical point,
and we have assumed that there are no degeneracies.
For degenerate systems like our experimental one, we just have to replace $|V_{01}|^2$ by
a sum over the degenerate subspace of the transition elements squared.
In a typical second order QPT, $\Delta \sim |\lambda-\lambda_c|^{-z\nu}$,
where $\nu$ is the correlation length critical exponent and $z$ is the dynamical
critical exponent.
As described in Sec. II, for a finite system the gap does not close but reaches a minimum
$\Delta_{min}$
that goes to zero with the size of the system $N$.
Thus, non-analyticities occur only in the thermodynamic
limit  $N \rightarrow \infty$.

\begin{figure}[bt]
\includegraphics[width=5in]{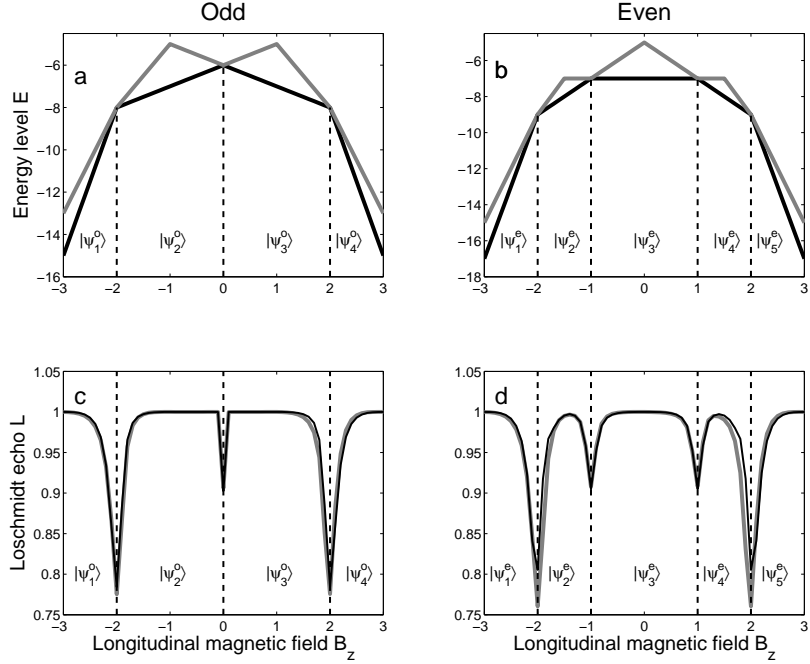}
%Matlab :phase.m and phase.fig in F:\UW\PhaseTrans
\caption{ Phase diagrams without transverse field (a-b) and
Loschmidt echo (dimensionless) with small transverse field (c-d)
for the Ising-chains with odd and even spins, shown in the left
and right columns, respectively. The dark and light curves in
figures (a-b) represent the two lowest energy levels, by setting
the coupling strength and $\hbar$ to unity. The phases and energy
levels are listed in Eqs. (\ref{odd}-\ref{energyeven}). The
crossover points are $B_c=\pm2$, $0$ in the odd spin system, and
$B_c=\pm2$, $\pm1$ in the even spin system. The minima of the
Loschomidt echo in panels (c) and (d) indicated the critical
points. Without loss of generality, we choose $N=7$ and $8$ to
illustrate the odd and even cases, where $\varepsilon=0.1$,
$\tau=\pi$, and $B_x=0.1$, for calculating $L$. In figures (c-d)
the light thick curves show the numerical results from Eq.
(\ref{define_L}), while the dark thin curves show the approximate
analytical results from Eq. (\ref{Lapprox}). }\label{fphase}
\end{figure}

Eq. (\ref{Lapprox}) suggests that whenever the ground and first excited states
are the most relevant for a particular system dynamics,
we can study the qualitative features of a QPT with a
two-level toy system under both transversal and longitudinal fields,
\be
H_{LZ}= \Delta_{min} \sigma_x +s( \lambda) |\lambda|^{z \nu} \sigma_z,
\label{LandauZener}
\ee
where $s(\lambda)$ is the sign function.
Furthermore, this toy model -- which represents Eq. (\ref{Lapprox}) exactly
up to ${\cal O}(\varepsilon^2)$ -- resembles the approximations we use to model a QPT
with our NMR quantum simulator -- see Fig. (\ref{fphase}) for a comparison between exact results
and this approximation.

From the spectra of our numerical simulations (see Fig. \ref{fphase}), we see that
our experiments are best described by $z\nu=1$. In this case, Eq. (\ref{LandauZener})
is the well known Landau-Zener model \cite{Landau} that has been
used successfully to predict the scaling laws for the creation of topological defects
when a system is quenched at finite speed through a critical point \cite{Damski05}.
For this Landau-Zener model,
\ba
\Delta=2\sqrt{\lambda^2+\Delta_{min}^2} \\
|V_{01}|^2 = \frac{\Delta_{min}^2}{\Delta_{min}^2+\lambda^2},
\label{rate}
\ea
which means it has a "critical point" at $\lambda=0$.
Expanding Eq. (\ref{Lapprox}) for short times, and replacing with Eq. (\ref{rate}),
\be
L(t)\simeq \exp\left(- \frac{\varepsilon^2 \Delta_{min}^2 t^2}
{\Delta_{min}^2+\lambda^2} \right).
%\exp(-\varepsilon^2 |V_{01}|^2 t^2).
\label{gaussian}
\ee
Since the decay rate of $L(t)$ (proportional to $|V_{01}|^2$)
has its maximum at $\lambda=\lambda_c=0$, then we
conclude that the decay of the echo is strongest
at the critical point -- or, conversely, that for a fixed time $t$ the
echo has a minimum at the critical point.

In order to discuss possible universal scaling properties of the
Loschmidt echo, our generalization in Eq. (\ref{LandauZener}) from a Landau-Zener model attempts to
incorporate a gap that closes with an arbitrary power $z\nu \ne1$.
In this general case the short time decay is still given by Eq. (\ref{gaussian}), with a
decay rate $\varepsilon^2 |V_{01}|^2$.
By choosing $V=\sigma_z$ independent of $\lambda$ and $\varepsilon$,
and taking $\Delta_{min}=1/N$ for demonstrative purposes, we find that
near the critical point,
\be
|V_{01}|^2 \underset{N\rightarrow\infty}{\sim} \frac{1}{N^2 |\lambda-\lambda_c|^{2 z\nu}}.
\ee
This suggests that the decay rate of the Loschmidt echo might show scaling with universal exponents.
Such scaling has been proven for the ground state fidelity and the quantum
geometric tensor \cite{Geometric}.
In principle, our experimental technique could be used to test universality and scaling properties
of the system.
However, our experiments are currently limited to the case $z\nu=1$ and relatively small sizes
that prevent us from exploring these properties.

\section{NMR implementation}

\subsection{Overview of the experiment}

\begin{figure}
\includegraphics[width=4in]{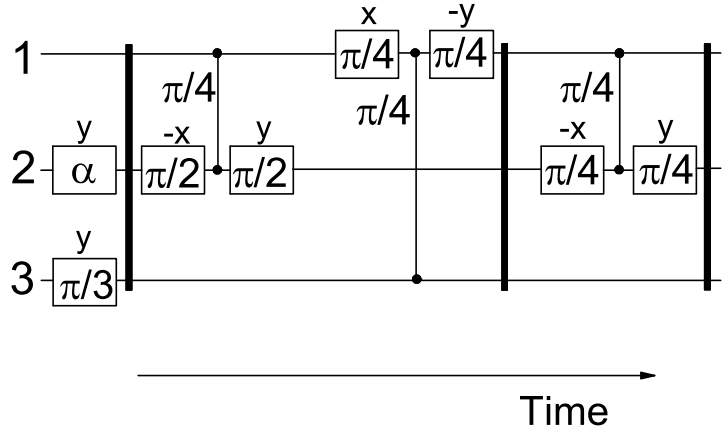}
%origin file: Pure4_s.opj in  F:\UW\PhaseTrans
\caption{ Gate sequence to prepare the effective pure state
$|000\rangle$ by spatial averaging from thermal equilibrium state
of TCE, where $\cos \alpha=2\gamma_{C}/\gamma_{H}$. Here
$\gamma_{H}$ and $\gamma_{C}$ denote the gyromagnetic ratios of
proton and carbon, respectively. The single qubit gates are
implemented through
 radio frequency pulses denoted by the rectangles. The rotation
angles and directions are shown inside and above the rectangles.
The bold vertical lines denotes the gradient pulses along $z$
axis. The two filled circles connected by a line denote the $J-$
coupling evolution $e^{-i\phi\sigma_{z}^{l}\sigma_{z}^{k}}$
between qubits $l$ and $k$, where $\phi$ is shown next to the
line. }\label{pulsepure3}
\end{figure}

\begin{figure}
\includegraphics[width=5in]{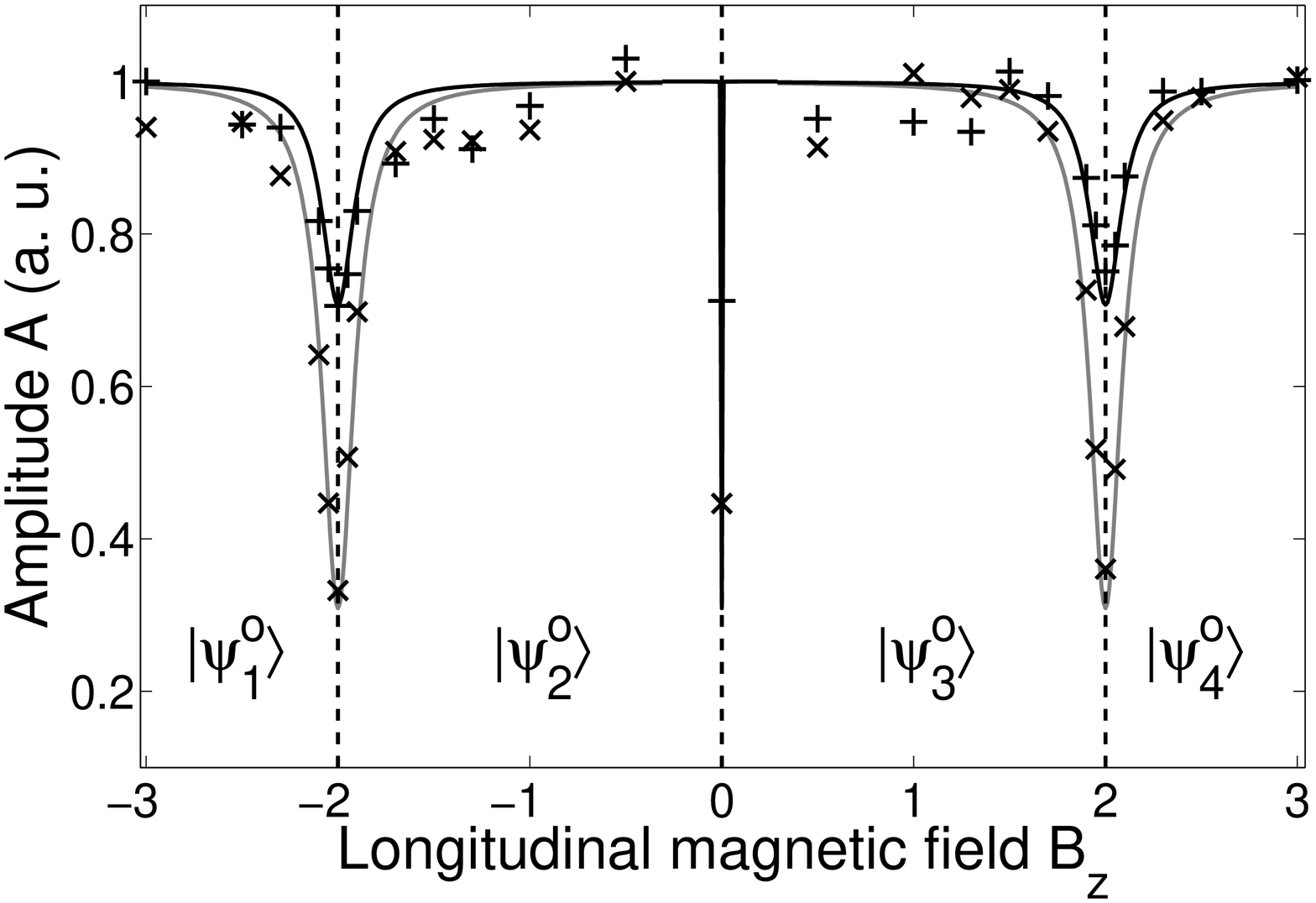}
%Matlab :Results_TCE.m and Results_TCE.fig in  F:\UW\PhaseTrans\LE3
\caption{Experimental results in the three qubit QPT system, where
$\tau=\pi$. The four phases $|\psi_{k}^{o}\rangle$ with $k=1$,
$\ldots$, $4$ are represented as
$|\psi_{k}^{o}\rangle=|000\rangle$, $|010\rangle$, $|101\rangle$,
and $|111\rangle$, respectively. The experimentally measured
amplitudes of the signals are marked by "$\times$" and $"+"$ for
$\varepsilon=0.2$ and $0.125$, respectively. The minima of the
amplitudes indicate the critical points. The theoretical results
are shown as the light and dark curves. The experimental results
show a good agreement with theory. }\label{results3}
\end{figure}

\begin{figure}
\includegraphics[width=5in]{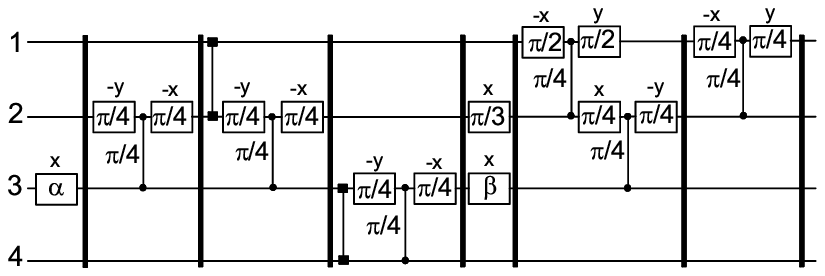}
%origin file: Pure4_s.opj in  F:\UW\PhaseTrans
\caption{ Gate sequence to prepare the effective pure state
$|0000\rangle$ by spatial averaging from the thermal equilibrium
state of the four carbons in crotonic acid, where
$\cos\alpha=1/8$, and $\cos\beta=1/4$. The filled rectangles in
pairs connected by a line denote a state specific swap gate
between qubits $l$ and $k$, i.e., it transforms $\sigma_z^{l}$ to
$\sigma_z^{k}$, or $\sigma_z^{k}$ to
$\sigma_z^{l}$.}\label{pulse4}
\end{figure}

\begin{figure}
\includegraphics[width=5in]{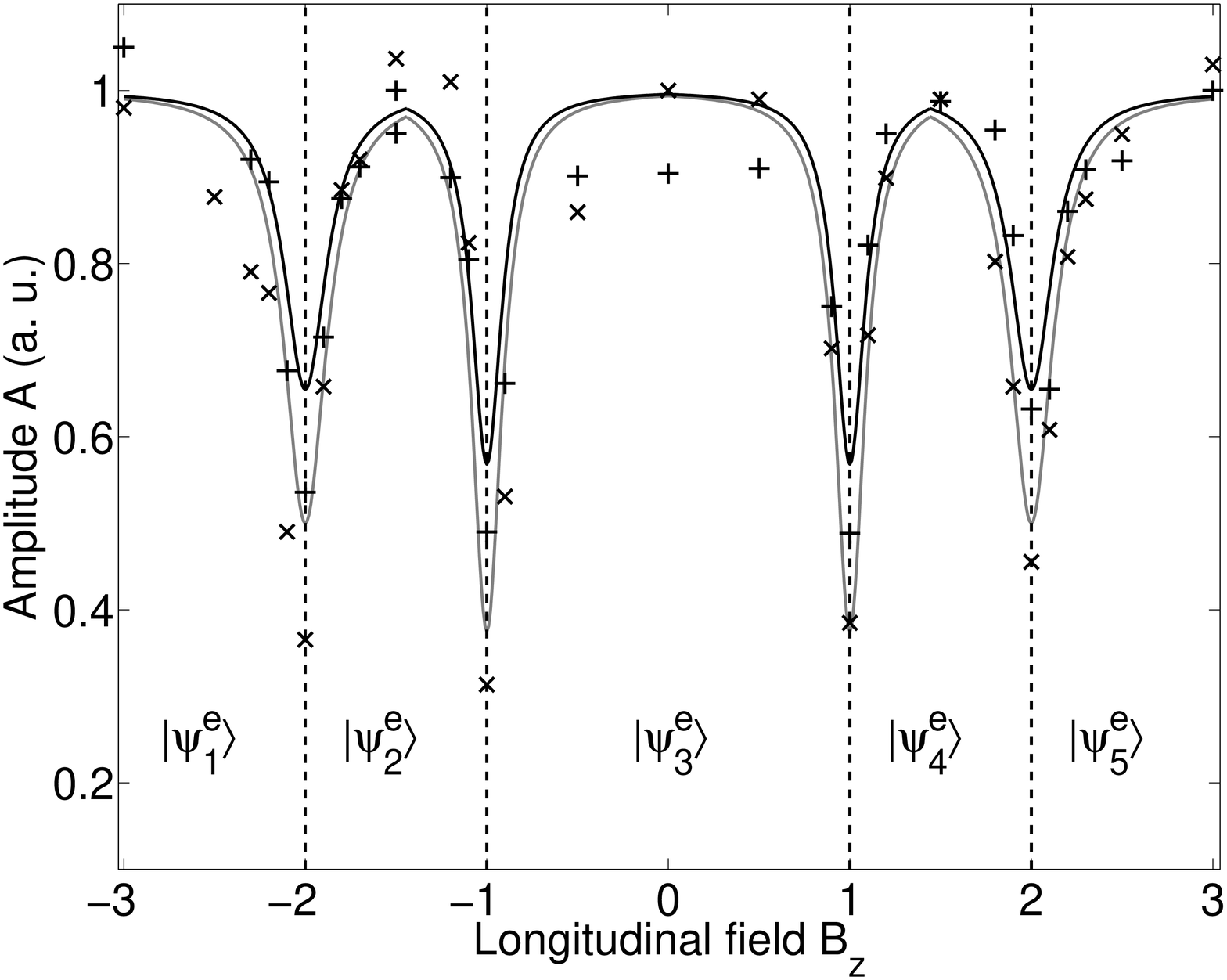}
%Results4.m in  F:\UW\PhaseTrans\LE4
\caption{Experimental results in the four qubit QPT system, where
$\tau=\pi/2$. The five phases $|\psi_{k}^{e}\rangle$ with $k=1$,
$\ldots$, $5$ are represented as
$|\psi_{k}^{e}\rangle=|0000\rangle$,
$(|0100\rangle+|0010\rangle)/\sqrt{2}$,
$(|0101\rangle+|1010\rangle)/\sqrt{2}$,
$(|1101\rangle+|1011\rangle)/\sqrt{2}$, and $|1111\rangle$,
respectively. The experimentally measured amplitudes
 are marked by
"$\times$" and "+" for $\varepsilon=0.5$ and $0.4$, respectively.
The minima of the amplitudes indicate the critical points. The
theoretical results are shown as the light and dark curves, in
good agreement with the experimental results. } \label{L4qubit}
\end{figure}

Our goal is to measure the Loschmidt echo in the antiferromagnetic
spin chain described by Hamiltonian (\ref{hamIsing}) as a
parameter ($B_z$) is varied, and from this infer the critical
points of the system. Step by step, the experiment can be
summarized as follows: Starting from the thermal equilibrium
state, we prepared a pseudo-pure state (gate sequences for this
are shown in Figures \ref{pulsepure3} and \ref{pulse4}). For each
value $B_z$ we transform the pseudo-pure state from the
computational basis to the ground state of the Hamiltonian
(\ref{hamIsing})  (that depends on $B_z$) using a unitary $U_0$.
We evolved the system forward in time with the Hamiltonian
(\ref{hamIsing}) at field $B_z$, and then backwards with a
perturbed field $B_z+\varepsilon$. After transforming the state
back to the computational basis using $U_0^{\dag}$, we encode $L$
as the diagonal element that is indicated by the initial
computational basis in the density matrix. Exploiting another
operation $D$ to eliminate the non-diagonal elements of the
density matrix, we can obtain the locations of the minima of $L$
using a selective readout pulse and observing the intensity of a
spectrum of a single qubit. We perform the experiment in chains of
three and four spins. The results are shown in Figures
\ref{results3} and \ref{L4qubit}, respectively.

\begin{figure}
\includegraphics[width=4in]{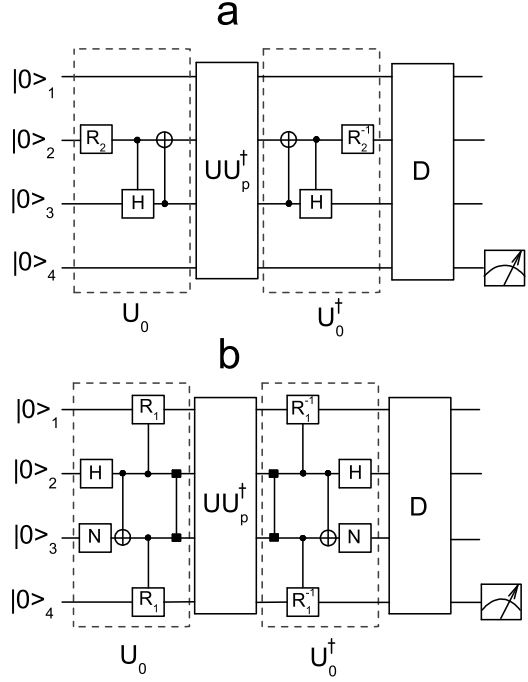}
%origin file: PulseSeq4_2.opj in  F:\UW\PhaseTrans
\caption{Quantum network for measuring critical points in
intervals $B_z\in [-3,-1.44]$, and $(-1.44,0]$ in four qubit
system, shown as figures (a-b) respectively. $H$ denotes the
Hadamard transform gate, and $U_p^\dagger U\approx
    e^{-i\tau\varepsilon(\sigma_{z}^{1}+\sigma_{z}^{2}+\sigma_{z}^{3}+\sigma_{z}^{4})}$.
     $R_2=e^{i\varphi_{2}\sigma_{y}}$ and
$R_1=e^{i\varphi_{1}\sigma_{y}}$, where $\varphi_2$ and
$\varphi_1$ are chosen as Eqs. (\ref{phi4m2}-\ref{phi4m1}). The
rectangle and the dot connected by a line denote a controlled
operation that is shown inside the rectangle. The filled
rectangles in pairs connected by a line denote a SWAP gate. The
networks for intervals $B_z \in [1.44,3]$ and $(0,1.44)$ can be
obtained by adding NOT gates to all qubits at the end of the
networks for implementing $U_0$ in figures (a-b), respectively. }
\label{ground4Net}
\end{figure}

We simplify the implementation of the experiment with a number of
approximations summarized here and described in detail in the
following sections. At each value of $B_z$ we prepare a very good
approximation of the ground state, with fidelity higher than $98\%$ % check by F3.m in \LE3, and F4 in \LE4
(we elaborate on this point in the conclusions). The approximated
ground state is obtained with perturbation theory around the
crossover point of zero transverse field and does not require
knowledge of the criticality of the system with non-zero
transverse field. We split the range of the field $B_z$ in
intervals (three and four for the odd and even spin chains
respectively) and use a different pulse sequence for each
interval. The forward-backward evolution is compressed into a
single step using a first order Trotter expansion with $98\%$
accuracy. The quantum networks for the odd and even chain
experiments are shown in Figures \ref{net3} and \ref{ground4Net}.
%After the evolution, we
%applied the inverse preparation gate $U_0^\dag$, and then
% another operation $D$ to eliminate the non-diagonal elements of the density
%matrix. The final gate transfers the Loschmidt echo signal into
%the amplitude of the initial pseudo-pure state, which we can read
%simply by measuring the spectrum of a single spin.

\subsection{Efficient detection of critical points using the Loschmidt echo}

In order to measure the Loschmidt echo we first prepare the ground
state $\ket{\psi(B_z,B_x)}$ of $H(B_z,B_x)$, which remains very
close to Eqs. (\ref{odd}) and (\ref{even}), except in the vicinity
of the critical points. Then, we evolve it forward under $H$ for a
period of time $t$, and next evolve it backwards under
$H+\varepsilon V$ for $t$, where $\varepsilon V$ is the fixed
perturbation introduced for detection with $|\varepsilon| \ll 1$.
Here, $B_z$ will be our control parameter ($\lambda$ in the
previous section), and we choose the perturbation as
$V=-\sum_{i=1}^{N}\sigma^{i}_{z}$. This choice of a global
perturbation simplifies our experiments, although more general
choices like local perturbations lead to the same results
but with a reduced signal
\cite{Rossini}. In order to detect the critical point of the
transition we fix the evolution time $t=\tau$ and the transversal
field $B_x$, and measure $L$ as a function of $B_z$
\cite{sunPRL06,Paz}. As shown in the previous section, the critical
points will be marked by the minima of
\begin{equation}\label{Ldefine}
L\equiv   \left. L(B_z)\right|_{t=\tau}=|\langle\psi(B_z,B_x)|U_{p}^\dagger U|\psi(B_z,B_x)\rangle|^{2},
\end{equation}
where $U=e^{-i\tau H}$ and $U_{p}=e^{-i\tau(H+\varepsilon V)}$ are
the unperturbed and perturbed evolution operators, respectively. We
show some representative echoes in small chains in Figure
\ref{fphase}(c-d).

Measuring an overlap such as Eq. (\ref{Ldefine}) in general might
require full state tomography techniques. Because of its
particular form, we can also couple the system to a probe qubit in
such a way  that $L$ is encoded in the off-diagonal terms of the
reduced density matrix of the probe \cite{Paz,Zhang08}. Here, we
present a method to measure $L$ directly in the system. We call
$U_{0}$  the unitary operation that prepares
$|\psi_{g}(B_z,B_x)\rangle$ from an arbitrary computation basis
state $|s\rangle$.
This is not necessarily an efficient operation for all systems
-- indeed, finding the ground state of arbitrary Hamiltonians might be
an NP-hard problem \cite{Poulin}. However, theoretical results suggest that any initial
state with a large overlap with the ground state is sufficient to detect
criticality \cite{Temperature}.
For instance, in our experiments we do not prepare
the true ground state of the system, but actually a state that
approximates it very well. We will discuss this and other alternatives
to the preparation of the ground state in the last section.

Through rewriting Eq. (\ref{Ldefine}) as
\begin{equation}\label{Lrewrite}
    L=|\langle s|U_{0}^{\dagger}U_{p}^{\dagger}U U_{0}|s\rangle|^{2},
\end{equation}
we find that $L$ can be obtained by projecting
\begin{equation}\label{Sfinal}
   |\Psi\rangle=U_{0}^{\dagger}U_{p}^{\dagger}U U_{0}|s\rangle
\end{equation}
onto state $|s\rangle$, i.e. $L$ is equal to the element
$|s\rangle\langle s|$ of the density matrix
$\rho=|\Psi\rangle\langle\Psi|$. Without loss of generality, we
chose $|s\rangle=|00...0\rangle$, the state with all qubits in
computational basis state $|0\rangle$. After the final evolution
$U_0^\dagger$, we eliminate the non-diagonal elements by gradient
pulses or dephasing processes \cite{Suter06,BellNMR}. Then,
through a read-out pulse, e.g. $\pi/2$, applied to an arbitrary
qubit, we obtain the signals marked by the states of other qubits.
%Gamble - good spot for a paragraph break

We are only concerned with the signal marked by the state in which
all other qubits are in state $|0\rangle$. Because in NMR we
observe differences in populations, the amplitude of this signal
$A$ is proportional to  $(L-\rho_{nn})\leq L$, with $n\neq 1$. The
locations of the minima of $A$ are the same as those of $L$, with
their values each decreased by an additional $\rho_{nn}$. This
allows us to detect the critical points through $A$ by measuring
only one qubit of the system.

\subsection{Odd N case}
We first demonstrate the detection of critical points of a QPT in
an odd spin system with $N=3$. We prepared an initial state that
approximates the ground state of the Hamiltonian for each value of
$B_z$. Using our notation for the ground states of $H$ for $B_x=0$
($|\psi_{k}^{o}\rangle=|000\rangle$, $|010\rangle$, $|101\rangle$,
and $|111\rangle$, for $k=1..4$ respectively), the ground state
near $B_c=\pm 2$ can be approximated as
\begin{equation}\label{groundm2}
 %   \psi_{g}(B_z,B_x)=|000\rangle\cos\varphi-|010\rangle\sin\varphi
\ket{\psi(B_z,B_x)}=|\psi_{m}^o\rangle\cos\varphi-|\psi_{n}^o\rangle\sin\varphi
\end{equation}
with
\begin{equation}\label{phim2}
%Gamble - modernized equation
\tan\varphi= \left[ (2-|B_z|)+\sqrt{(2-|B_z|)^2+B_x^2} \right]/B_x,
\end{equation}
where $m=1$, $n=2$ or $m=4$, $n=3$, corresponding to $B_c=-2$ or
$2$, respectively. In the vicinity of $B_c=0$, the gap between the
lowest energy levels is so small that the ground state can be well
approximated by $|\psi_{2}^{o}\rangle$, $(|\psi_{2}^{o}\rangle-
|\psi_{3}^{o}\rangle)/\sqrt{2}$, or $|\psi_{3}^{o}\rangle$,
 corresponding to $B_z<0$, $B_z=0$, or
$B_z>0$, respectively. %checked by Heff3.m and chain3_echo.m in \...\LE3
%the effective Hamiltonian is
%represented as
%\begin{equation}\label{Heff0}  %checked by Heff3.m and chain3_echo.m
%    H_{0}=\left ( \begin{array}{cccc}
%                    B_z-2 & 0 & \sqrt{2}B_x & 0 \\
%                    0 & B_z & B_x & \sqrt{2}B_x \\
%                    \sqrt{2}B_x & B_x & -B_z & 0 \\
%                    0 & \sqrt{2}B_x & 0 & -B_z-2
%                  \end{array}
%            \right ).
%\end{equation}
%Hence we obtain the effective Hamiltonian in space $\{
%|010\rangle, |101\rangle\}$
%\begin{equation}\label{Heff0app}
%    H_{0}=-2B_z I+B_z\sigma_z.
%\end{equation}

  For the experimental implementation, we used $^{13}$C
labelled trichloroethylene (TCE), dissolved in d-chloroform as the
sample \cite{tce}. Data were taken with a Bruker DRX 700 MHz
spectrometer. We denote the $^{1}$H nuclear spin as qubit 2 (H2),
the $^{13}$C directly connected to $^{1}$H is denoted as qubit 1
(C1), and the other $^{13}$C as qubit 3 (C3). The difference of
frequency between C1 and C3 is about $1249.2$ Hz, and the coupling
constants are $J_{13}=103.1$Hz, $J_{12}=200.9$Hz, and
$J_{23}=9.16$Hz. The spin-selective excitation for C1 or C3 is
realized by a GRAPE pulse \cite{grape}. The $J$-coupling
evolution $e^{-i\phi\sigma_{z}^{l}\sigma_{z}^{k}}$ between qubits
$l$ and $k$ is implemented by a standard refocusing pulse
sequence \cite{refocus}.
%Because of the strongly
%coupled carbons \cite{Miquel} we describe the Hamiltonian of the
%three-qubit system as
%\begin{equation}\label{HTCE}
%  H=-\pi\sum_{i=1}^3 \nu_{i}\sigma_{z}^{i}
%  +\frac{\pi}{2} J_{12}\sigma_{z}^{1}\sigma_{z}^{2}
%  +\frac{\pi}{2} J_{23}\sigma_{z}^{2}\sigma_{z}^{3}
%  +\frac{\pi}{2} J_{13}(\sigma_{x}^{1}\sigma_{x}^{3}
%  +\sigma_{y}^{1}\sigma_{y}^{3}
%  +\sigma_{z}^{1}\sigma_{z}^{3}).
%\end{equation}
The effective pure state $|000\rangle$ is prepared by spatial
averaging \cite{effectivePure} from the thermal equilibrium state
$\rho_{eq}=\gamma_{H}\sigma_{z}^{2}+
  \gamma_{C}(\sigma_{z}^{1}+\sigma_{z}^{3})$, by approximating the system
as a weakly-coupling system,  where $\gamma_{H}$ and $\gamma_{C}$
denote the gyromagnetic ratios of proton and carbon, respectively.
The gate sequence for the pseudo-pure state preparation is shown as Figure \ref{pulsepure3}.
%The following radio-frequency (rf) and magnetic
%field gradient pulse sequence transforms the system from the
%thermal equilibrium state
%\begin{equation}\label{equ}
%  \rho_{eq}=\gamma_{H}\sigma_{z}^{2}+
%  \gamma_{C}(\sigma_{z}^{1}+\sigma_{z}^{3})
%\end{equation}
%to $|000\rangle$:
%\begin{eqnarray}
%% \nonumber to remove numbering (before each equation)
%&&[\phi]_{y}^{2}-[\frac{\pi}{3}]_{y}^{3}-[grad]_{z}\nonumber\\
%&-&[-\frac{\pi}{2}]^{2}_{x}-[\frac{1}{2J_{12}}]-
%[\frac{\pi}{2}]^{2}_{y}-[\frac{\pi}{4}]^{1}_{x}-[\frac{1}{2J_{13}}]-[-\frac{\pi}{4}]^{1}_{y}-[grad]_{z} \nonumber\\
%&-&[-\frac{\pi}{4}]^{2}_{x}-[\frac{1}{2J_{12}}]-[\frac{\pi}{4}]^{2}_{y}-[grad]_{z}.
%\end{eqnarray}
% Here $\gamma_{H}$, and $\gamma_{C}$ denote the gyromagnetic ratios of
%proton, and carbon, respectively, and $\cos
%\phi=2\gamma_{C}/\gamma_{H}$. $[grad]_{z}$ denotes a gradient
%pulse along the $z$- axis. $[\pi/2]_{x}^{1}$ denotes a $\pi/2$
%pulse along the $x$- axis acting on the H2 qubit.

\begin{figure}
\includegraphics[width=5in]{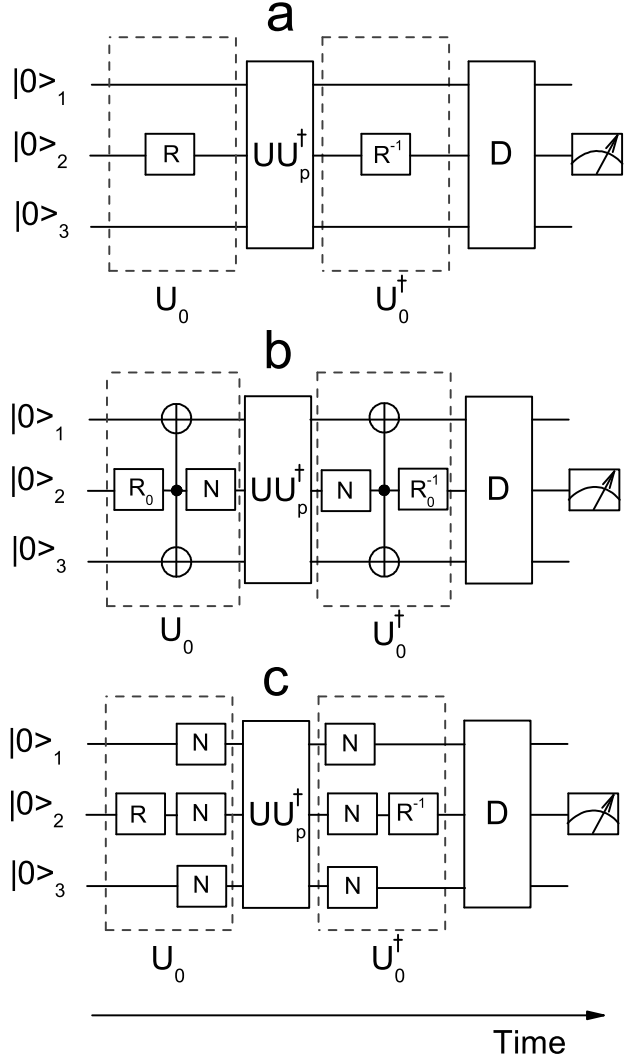}
%Matlab :Network3.ojp as an origin file in F:\UW\PhaseTrans
%ref QPT3.seq and QPTp2.seq in F:\UW\PhaseTrans\LE3
\caption{Quantum networks for measuring critical points in
intervals $B_z\in [-3,-1]$, $(-1,1)$, and $[1,3]$ in the three
qubit system, shown as figures (a-c) respectively.
$R=e^{i\varphi\sigma_{y}}$ where $\varphi$ is given by Eq.
(\ref{phim2}), and $R_0=\mathbf{1}$ (unit operator),
$e^{i\pi\sigma_{y}/4}$ or $e^{i\pi\sigma_{y}/2}$ for $B_z=-0.5$,
$0$, or $0.5$, respectively. $U_0$ and $U_0^\dag$ are indicated by
the dashed rectangles, and $U_p^\dagger U\approx
    e^{-i\tau\varepsilon(\sigma_{z}^{1}+\sigma_{z}^{2}+\sigma_{z}^{3})}$. $\bigoplus$ and the
black dot connected by a line denote a controlled NOT gate, and
 $N$ denotes a NOT gate. $D$ denotes the operation to eliminate the non-diagonal
elements of the density matrix. The last operation  in each figure
denotes the measurement, which can be applied to an arbitrary
qubit of the system. }\label{net3}
\end{figure}

\begin{figure}
\includegraphics[width=5in]{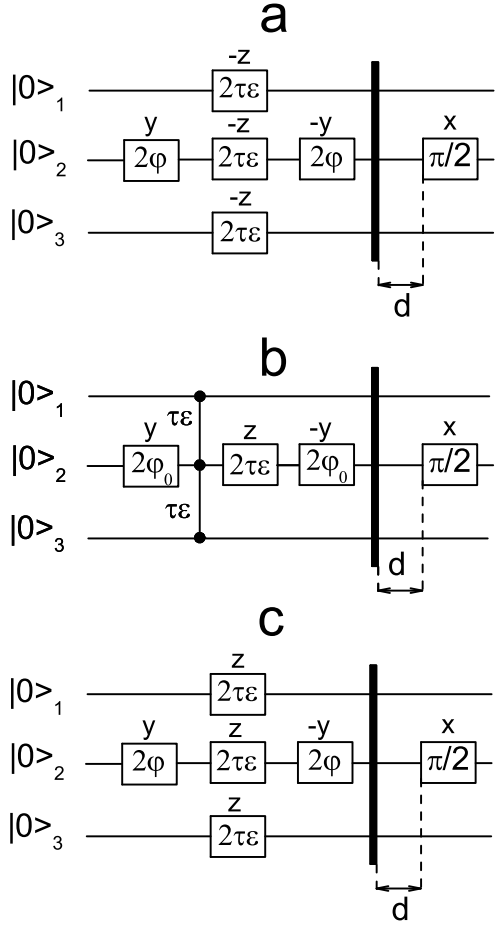}
%Matlab :PulseSeq3.ojp as an origin file in F:\UW\PhaseTrans
%ref QPT3.seq and QPTp2.seq in F:\UW\PhaseTrans\LE3
%check.m to check Z1 commutes with CNOT_12
\caption{Gate sequences (a-c) to implement Figures \ref{net3}
(a-c), respectively.
%Gamble - this next part could be more clear. The only obvious circles are open, in figure a. Also, I would say "bold vertical line" rather than high narrow filled rectangle.
A 16-step average over a random delay, denoted by $d$, between 0
and 10 ms, dephases the residual zero-quantum coherence. The last
$\pi/2$ pulse is the readout pulse, which can be applied to an
arbitrary qubit of the system.}\label{preground}
\end{figure}

 In order to measure the echo we split the $B_z$ axis in intervals near
  $B_c=-2$, $0$, and $2$. In particular, we use different
quantum networks for $B_z\in [-3,-1]$, $(-1,1)$, and $[1,3]$, show
in Figures \ref{net3} (a-c) respectively. The operations for
preparing $U_0$ and $U_0^{\dag}$ are indicated by the dashed
rectangles and $D$ denotes the operation to eliminate the
non-diagonal elements of the density matrix. Figure
\ref{preground} shows the corresponding gate sequences.  The
evolution time is chosen as $\tau=\pi$, and $\varepsilon=0.2$ or
$0.125$. The echo evolution $U_p^\dagger U$ can be approximated by
%\begin{equation}\label{Uexp}
  $  U_p^\dagger U\approx
    e^{-i\tau\varepsilon(\sigma_{z}^{1}+\sigma_{z}^{2}+\sigma_{z}^{3})}$
%\end{equation}
%and implemented by
%\begin{equation}\label{echo3}
%    [\frac{\pi}{2}]_{y}^{1,2,3}-[2t\varepsilon]_{x}^{1,2,3}-[-\frac{\pi}{2}]_{y}^{1,2,3}.
%\end{equation} %seeing eq.(26) in PHYSICAL REVIEW A 73, 062325 (2006) 'speedup...'
with fidelity larger than $98\%$. We optimize the gate sequence
$CNOT_{21}- e^{-i\tau\varepsilon\sigma_{z}^{1}}-CNOT_{21}$ as
$e^{-i\tau\varepsilon\sigma_{z}^{1}\sigma_{z}^{2}}$, and
$CNOT_{23}- e^{-i\tau\varepsilon\sigma_{z}^{3}}-CNOT_{23}$ as
$e^{-i\tau\varepsilon\sigma_{z}^{2}\sigma_{z}^{3}}$
\cite{zhangpra04} to obtain figure \ref{preground} (b) from Figure
\ref{net3} (b). The amplitudes of signals are obtained by
measuring on H2, with experimental results shown in Figure
\ref{results3}. Experimental data are marked by "$\times$" and "+"
for $\varepsilon=0.2$ and $\varepsilon=0.125$, respectively. The
corresponding theoretical results are indicated by the light
 and dark  curves. The critical points are
correctly indicated by the minima of the amplitudes.

\subsection{Even N case}

   We illustrate the detection of QPT critical points in an even spin chain with
   $N=4$. Here we use the notation for the ground states for $B_x=0$,
$|\psi_{k}^{e}\rangle=|0000\rangle$,
$(|0100\rangle+|0010\rangle)/\sqrt{2}$,
$(|0101\rangle+|1010\rangle)/\sqrt{2}$,
$(|1101\rangle+|1011\rangle)/\sqrt{2}$, and $|1111\rangle$,
for $k=1..5$ respectively.
Depending on the value of $B_z$ we prepare an approximation to the ground state
\begin{equation}\label{gstate4}
\psi(B_z,B_x)=|\psi_{m}^e\rangle\cos\varphi-|\psi_{n}^e\rangle\sin\varphi.
\end{equation}
For $B_z$ near $B_c=\pm2$
\begin{equation}\label{phi4m2} %check by chain4_theory.m in \PhaseTrans\L4
\tan\varphi=\ [ (2-|B_z|)+\sqrt{(2-|B_z|)^2+2B_x^2}\
]/(\sqrt{2}B_x)
\end{equation}
with $m=1$, $n=2$, or $m=5$, $n=4$, corresponding to $B_c=-2$ or
$2$ respectively. For $B_z$ near $B_c=\pm1$ we use
\begin{equation}\label{phi4m1}
\tan\varphi=\ [(1-|B_z|)+\sqrt{(1-|B_z|)^2+B_{x}^{2}}\ ]/B_{x}
\end{equation}
with  $m=2$, $n=3$ or $m=4$, $n=3$, corresponding to $B_c=-1$ or
$1$, respectively.

  For implementation, we choose the four carbons in crotonic acid
  \cite{crot} dissolved in d6-acetone as the four qubits by decoupling the protons.
Data were taken with a Bruker DRX 700 MHz spectrometer. The
chemical shifts for the four carbons $\nu_{1-4}$ are $-2965.75$,
$-25501.9$, $-21583.9$ and $-29431.5$ Hz. The $J$-couplings are
$J_{12}=41.6$, $J_{23}=69.7$, $J_{34}=72.0$, $J_{13}=1.5$,
$J_{14}=7.0$, and $J_{24}=1.2$ Hz.

   We prepare the pseudo-pure state
by spatial averaging through improving the scheme found in
\cite{CoryPure4}. Our technique can be illustrated by transforming
the thermal equilibrium  state of a four qubit system
$\sum_{i=1}^{4}\sigma_{z}^{i}$ to
\begin{equation}
\left(\sum_{i=1}^{3}\sigma_{z}^{i}\right)(\mathbf{1}/2+\sigma_{z}^{4})+\sigma_{z}^{4}/8,
\end{equation}
where $\mathbf{1}$ denotes the unit operator
 and $\sum_{i=1}^{3}\sigma_{z}^{i}$ can be transformed to an effective
pure state in the three-qubit system. This method generalizes to
an $N$-qubit system in a recursive manner. After some
simplifications \cite{weiCP}, the complete gate sequence to
generate $|0000\rangle$ is shown as Figure \ref{pulse4}, where the
state specific swap gate requires two $J-$couplings with evolution
time $1/(2J_{lk})$ \cite{swap}. In the ideal case the strength of
the single peak obtained through a $\pi/2$ read out pulse
selective for one spin is equal to that of the same peak  in a
spectrum of  the thermal state, where eight peaks with equal
strength appear.
%Compared
%with the previous scheme, we use less $J-$ couplings, and do not
%use the long distance coupling.

\begin{figure}
\includegraphics[width=5in]{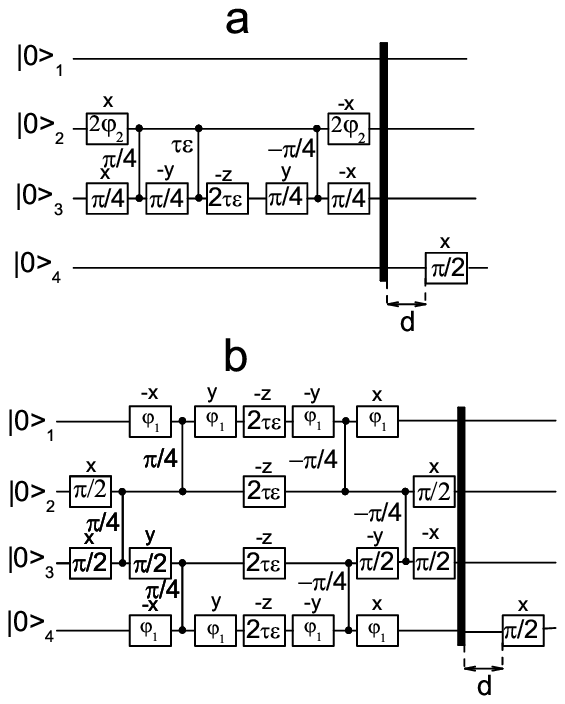}
%origin file: Ground.opj in  F:\UW\PhaseTrans
\caption{Quantum gate sequences (a-b) to implement Figures
\ref{ground4Net} (a-b), respectively. Through replacing
$\varepsilon$ by $-\varepsilon$ in figure (a-b) one can obtain the
gate sequences for the intervals  $B_z \in [1.44,3]$ and
$(0,1.44)$, respectively.} \label{ground4}
\end{figure}

  The ground states are prepared from Eqs. (\ref{gstate4} -
\ref{phi4m1}). As before, we split the $B_z$ axis in intervals
around the critical points of zero transverse field. The networks
to measure the echo for $B_z\in [-3,-1.44]$, and $(-1.44,0]$ are
shown in Fig. \ref{ground4Net} (a-b). From these one can obtain
the networks for the intervals $B_z \in [1.44,3]$ and $(0,1.44)$
simply by adding NOT gates to all qubits at the end of the
corresponding networks for implementing $U_0$. Through compiling
the pulse sequence \cite{compile}, we obtain the gate sequences
shown as Figure \ref{ground4}, where
  $U_p^\dagger U\approx
    e^{-i\tau\varepsilon(\sigma_{z}^{1}+\sigma_{z}^{2}+\sigma_{z}^{3}+\sigma_{z}^{4})}$
with fidelity larger than $98\%$ and the two SWAP gates are
cancelled because they commute with $e^{-i\tau\varepsilon
(\sigma_{z}^{2}+\sigma_{z}^{3})}$.
%check by and chain4_thoery.m for fidelity, and  SWAP.m in PhaseTrans
 Experimental results are shown
in Figure \ref{L4qubit}, with $\tau=\pi/2$. The measured
amplitudes are marked by "$\times$" and "+" for $\varepsilon=0.5$
and $0.4$, respectively. The solid curves shows the corresponding
theoretical results. Again, the critical points are correctly
indicated by the minima of the amplitudes, so the experiment
results are in good agreement with theoretical expectations. The
observed errors could be explained by imperfections in the
implementation of the radio frequency pulses, inhomogeneities of
magnetic fields and decoherence.

\section{Discussion and Conclusions}

  We performed experimentally  quantum simulations of the second
order quantum phase transitions in finite systems. In particular,
we showed the QPTs and found the critical points of three- and
four-spin Ising chains, representative of odd and even spin
chains, respectively.
 The critical points are indicated by the
minima of the Loschmidt echo. We showed that this echo can be
realized by inducing the perturbation with an external field, and
the positions of its minima (related to the critical points) can
be obtained by measuring only an arbitrary qubit of the system. In
the weakly and fully resolved coupling systems, the resonance
lines can be assigned, and the line marked by the other qubits in
$|0\rangle$ can be identified. However, in large size systems
where the requirement of fully resolved couplings is not
practical, or in the strongly coupling systems, e.g., liquid
crystal or solid NMR systems, where the assignment of resonance
lines are not possible, one cannot identify the marked line. For
these cases, our method can be generalized by measuring the global
polarizations of the whole system by a collective $\pi/2$ pulse
(or $N$ pulses selective for each qubit), replacing the readout
pulse applied to one qubit. In the vicinities of the critical
points, the loss of the polarization due to the decoherence
process (e.g., gradient pulse or dephasing process) approaches the
maxima. Hence the critical points will be indicated by the minima of
the amplitude of the total signals of all qubits. Furthermore, this
has the advantage that a
global measurement is scalable with the size of the system.
 %check by J_resovled2.m J_resovled2.m in \LE4 and J_resolved3.m in \LE3
 %and LE_LC.m in \LXSTAL for liquid crystal system

 Our method improves the previous
one that required a probe qubit for both the perturbation and
the measurement \cite{Zhang08}. We believe this advantage gives
our method better scalability with the size of the system. In
particular, the perturbations created by the probe qubit method
are limited by the probe-system coupling strength, and,
furthermore, can become weaker than the noise in large systems
when they do not couple the probe to a macroscopic number of
normal modes in the system. Separating the perturbation and
measurement also gives finer control over the whole experiment.

On the issue of scalability, a very important point in the
algorithm is the preparation of the initial state. From a theory
point of view, most of the studies of the Loschmidt echo have used
ground states as initial states only because of simplicity.
However, preparing the ground state of an arbitrary Hamiltonian is
an NP-hard problem \cite{Poulin}. Furthermore, it would be
redundant, since it is most likely that knowing the exact ground
states is equivalent to knowing everything about the system --
including the information about criticality that one wants to
obtain from the echo experiments. Nonetheless, there is evidence
that the initial state need not be the exact ground state, but any
state with a sizeable overlap with the ground state. For instance,
analytical studies show that thermal states at temperatures at or
below the energy scales of the system can be used effectively to
detect the quantum phase transition \cite{Temperature}, where the
number of the spins can be up to 200. However, in some systems
(like our liquid NMR experiments) preparing a thermal state is not
particularly easier than other --perhaps more useful-- states. For
instance, in our experiments we prepared a good approximation to
the ground state that we obtained from a simple perturbation
theory around the crossover point of zero transverse field.
%The scalability of our method can be
%illustrated as following. It is sufficient to discuss the
%criticality around $B_c=\pm 2$, because when $N$ is infinite the
%"quasi"- phases around $B_c=0$ for odd $N$, or $\pm 1$ for even
%$N$, merge into a single antiferromagnetic phase. Without loss of
%generality we prepare the ground state around $B_c=-2$ for the
%system of small transverse field, and assume $N$ is an odd number.
%By choosing states $\{|\phi_k\rangle\}$ with $k=1$, $\ldots$,
%$(N+1)/2$,  represented as
%\begin{equation}\label{basisH1}
%    |\phi_{1}\rangle=|\underbrace{00...0}_{N}\rangle
%\end{equation}
%\begin{equation}\label{basisHk}
%    |\phi_{k}\rangle=(|\underbrace{01...01}_{k-1 \hspace{0.1cm} pairs \hspace{0.1cm} of \hspace{0.1cm}
%    01 \hspace{0.1cm}
%    }\underbrace{00...0}_{N-2(k-1)}\rangle
%    +|\underbrace{00...0}_{N-2(k-1)\hspace{0.2cm}}\underbrace{10...10}_{k-1 \hspace{0.1cm} pairs \hspace{0.1cm} of \hspace{0.1cm} 10}\rangle)/\sqrt{2}
%\end{equation}
%for $k=2$, $\ldots$, $(N-1)/2$ and
%\begin{equation}\label{basisHN}
%    |\phi_{\frac{N+1}{2}}\rangle=|\underbrace{01...01}_{(N-1)/2 \hspace{0.1cm} pairs \hspace{0.1cm} of \hspace{0.1cm} 01
%    }0\rangle
%\end{equation}
%as the basis, one can obtain the effective Hamiltonian $H_{eff}$
%with $(N+1)/2$ dimension. The numerical results show that the
%ground state of $H_{eff}$ can work well for detecting the critical
%point using the Loschmidt echo. The dimension of $H_{eff}$
%increases linearly with $N$. Therefore our method is scalable with
%the size of the system.
This method suggests that other
approximations, such as mean field or numerical classical
algorithms, could work well to detect criticality.

While the problem of finding strict
minimum requirements for the initial state of the algorithm
is clearly in need of more research,
we feel that it is reasonable to argue
that initialization of the algorithm is scalable: it only requires
finding among many possibilities one that can be prepared
efficiently in a quantum computer.  It would be
interesting to study the effect of more efficiently-prepared ground
states \cite{Poulin} or to investigate if state-independent
indicators -- such as the operator fidelity susceptibility
proposed in Ref. \cite{Wang08} -- could get rid of the initial
state issue altogheter. Finally, we would like to mention that
other possible extensions of our experimental methods are using
the Loschmidth echo to measure QPTs in gapless systems
\cite{Yang07,Wang08}, and also for measuring thermal phase
transitions \cite{You,Quanpreparation}.

\section{Acknowledgements}
We thank D. Suter, J. P. Paz, C. Batista, G. Ortiz, T.-C. Wei, M.
Zwolak, and H.T. Quan for helpful discussions. F.M.C. acknowledges
support from DOE-LDRD programs. J.K.G. acknowledges support from
the NSF-REU and Los Alamos Summer School programs.

\appendix
\section{Perturbative expansion of the Loschmidt echo}\label{appendix}

We start from the expansion of Eq. (\ref{taylor}),
\be
\ell(t)\simeq
 \left. \ell(t)\right|_{\varepsilon=0}
 +
 \left. \frac{\partial \ell(t)}{\partial \varepsilon} \right|_{\varepsilon=0}
 \varepsilon
+
\left. \frac{\partial^2 \ell(t)}{\partial \varepsilon^2} \right|_{\varepsilon=0}
\frac{\varepsilon^2}{2},
\label{taylorA}
\ee
where
\be
\ell(t)=\bra{0(\lambda)}e^{i H_0 t}e^{-i (H_0+\varepsilon V) t}  \ket{0(\lambda)},
\ee
with $\ket{0(\lambda)}$ the ground state of $H_0$, and we keep  the harmless $e^{i H_0 t}$
operator because it will simplify the results.
The first term of the expansion can be simply evaluated as in Eq. (\ref{mu0}),
\be
 \left. \ell(t)\right|_{\varepsilon=0} = \bra{0(\lambda)} e^{i H_0 t}e^{-i H_0 t} \ket{0(\lambda)} =1.
\label{mu0A}
\ee
For the first and second order terms we must compute derivatives of the evolution operator.
We can do this by expanding the exponential into an infinite series sum,
\ba
 \left. \frac{\partial \ell(t)}{\partial \varepsilon} \right|_{\varepsilon=0}
& =& \bra{0(\lambda)} e^{i H_0 t}
\left. \frac{\partial  e^{-i (H_0+\varepsilon V) t}}{\partial \varepsilon} \right|_{\varepsilon=0}
\ket{0(\lambda)} \nonumber \\
&  =  &
\bra{0(\lambda)} e^{i H_0 t}
\frac{\partial }{\partial \varepsilon} \sum_{n=0}^{\infty} \left. \frac{1}{n!}(-i (H_0+ \varepsilon V) t)^n \right|_{\varepsilon =0}  \ket{0(\lambda)}
\nonumber \\
& = &\bra{0(\lambda)}e^{i H_0 t} \sum_{n=1}^{\infty} \frac{(-it)^n}{n!} \sum_{k=0}^{n-1} \left. (H_0+ \varepsilon V)^k V (H_0+ \varepsilon V)^{n-1-k} \right|_{\varepsilon =0}  \ket{0(\lambda)}  \nonumber \\
& = & \bra{0(\lambda)}
e^{i H_0 t}
\sum_{n=1}^{\infty} \frac{(-it)^n}{n!} \sum_{k=0}^{n-1} H_0^k V H_0^{n-1-k} \ket{0(\lambda)}.
\label{firstderivative}
\ea
Computing now the expectation value,
\ba
 \left. \frac{\partial \ell(t)}{\partial \varepsilon} \right|_{\varepsilon=0}
& =& e^{i E_0 t} \sum_{n=1}^{\infty} \frac{(-it)^n}{n!} \sum_{k=0}^{n-1} E_0^k V E_0^{n-1-k}  \nonumber \\
& = & e^{i E_0 t} \sum_{n=1}^{\infty} \frac{(-it)^n}{n!} E_0^{n-1} n \bra{g(\lambda)} V \ket{g(\lambda)} \nonumber \\
& = & (-i t) V_{00} e^{i E_0 t} \sum_{m=0}^\infty \frac{(-itE_0)^m}{m!} \nonumber \\
& = & (-i t) V_{00},
\label{mu1}
\ea
where $E_0$ is the ground state energy and
$V_{00}=\bra{0(\lambda)} V \ket{0(\lambda)}$.

For the second order term we continue derivating Eq. (\ref{firstderivative}) before the evaluation
at $\varepsilon=0$,
\ba
 \left. \frac{\partial^2 \ell(t)}{\partial \varepsilon^2} \right|_{\varepsilon=0}
&=& \bra{0(\lambda)} e^{i H_0 t}
\left. \frac{\partial^2  e^{-i (H_0+\varepsilon V) t}}{\partial \varepsilon^2} \right|_{\varepsilon=0}
\ket{0(\lambda)} \nonumber \\
& = &
\bra{0(\lambda)} e^{i H_0 t}
\frac{\partial }{\partial \varepsilon}
\sum_{n=1}^{\infty} \frac{(-it)^n}{n!} \sum_{k=0}^{n-1} \left. (H_0+ \varepsilon V)^k V (H_0+ \varepsilon V)^{n-1-k} \right|_{\varepsilon =0} \ket{0(\lambda)} \nonumber \\
& = & \bra{0(\lambda)} e^{i H_0 t}
\sum_{n=2}^{\infty} \frac{(-it)^n}{n!} \left[
\left(
\sum_{k=1}^{n-1}  \sum_{m=0}^{k-1} (H_0+ \varepsilon V)^m V (H_0+ \varepsilon V)^{k-1-m}
V (H_0+ \varepsilon V)^{n-1-k}
\right)  \right.  \nonumber \\
& & +
\left. \left(
\sum_{k=0}^{n-2}  (H_0+ \varepsilon V)^k V \sum_{m=0}^{n-2-k} (H_0+ \varepsilon V)^m V (H_0+ \varepsilon V)^{n-2-k-m}
\right)
\right]_{\varepsilon =0}\ket{0(\lambda)} \nonumber \\
& = &
\bra{0(\lambda)} e^{i H_0 t}
\sum_{n=2}^{\infty} \frac{(-it)^n}{n!} \left[
\left(
\sum_{k=1}^{n-1}  \sum_{m=0}^{k-1} H_0^m V H_0^{k-1-m}
V H_0^{n-1-k}
\right)  \right. \nonumber \\
&  &+
\left. \left(
\sum_{k=0}^{n-2}  H_0^k V \sum_{m=0}^{n-2-k} H_0^m V H_0^{n-2-k-m}
\right)
\right] \ket{0(\lambda)}.
\label{secondderivative}
\ea
By taking the expectation value on the ground state we now obtain
\ba
\left. \frac{\partial^2 \ell(t)}{\partial \varepsilon^2} \right|_{\varepsilon=0}
& = & e^{i E_0 t} \sum_{n=2}^{\infty} \frac{(-it)^n}{n!} \left[
\left(
\sum_{k=1}^{n-1}  \sum_{m=0}^{k-1} E_0^{n-1-k+m}
\bra{0(\lambda)} V H_0^{k-1-m} V \ket{0(\lambda)}
\right)  \right. \nonumber \\
&  &+
\left. \left(
\sum_{k=0}^{n-2}  \sum_{m=0}^{n-2-k}  E_0^{n-2-m}
\bra{0(\lambda)}  V H_0^m V \ket{0(\lambda)}
\right)
\right],
\ea
replacing now $k'= k-1$ and $m'=m+k$ in the first and second sums inside the brackets
\ba
\left. \frac{\partial^2 \ell(t)}{\partial \varepsilon^2} \right|_{\varepsilon=0}&=&
e^{i E_0 t}
\sum_{n=2}^{\infty} \frac{(-it)^n}{n!}
E_0^{n-2}
\left[
\left(
\sum_{k=0}^{n-2}  \sum_{m=0}^{k}
E_0^{-k+m}
\bra{0(\lambda)} V H_0^{k-m} V \ket{0(\lambda)}
\right)  \right. \nonumber \\
&  & +
\left. \left(
\sum_{k=0}^{n-2}  \sum_{m=k}^{n-2}
E_0^{-m+k}
\bra{0(\lambda)}  V H_0^{m-k} V \ket{0(\lambda)}
\right)
\right] \nonumber \\
& = & e^{i E_0 t}
\sum_{n=2}^{\infty} \frac{(-it)^n}{n!}
E_0^{n-2}
\left[ (n-1) \bra{0(\lambda)} V^2 \ket{0(\lambda)}
\right. \nonumber \\
& & + \left.
\sum_{k=0}^{n-2}
\sum_{m=0}^{n-2}
E_0^{-|k-m|}
\bra{0(\lambda)} V H_0^{|k-m|} V \ket{0(\lambda)}
\right].
\ea
We can simplify the term inside the brackets by counting the number of times the terms
with $|k-m|=0$, $|k-m|=1$, and so on are repeated. The final expression is then
\ba
\left. \frac{\partial^2 \ell(t)}{\partial \varepsilon^2} \right|_{\varepsilon=0}&=&
e^{i E_0 t}
 \sum_{n=2}^{\infty} \frac{(-it)^n}{n!}
E_0^{n-2}
\left[2
\sum_{k=0}^{n-2}
E_0^{-k} (n-1-k)
\bra{0(\lambda)} V H_0^{k} V \ket{0(\lambda)}
\right].
\ea
We can make further progress by inserting identities $\sum_{\alpha=0}^{N-1} \ket{\alpha}\bra{\alpha}$,
with $\ket{\alpha}$ the basis of eigenstates of $H_0$ (we assume a finite Hilbert space $\alpha=0,..,N-1$),
\ba
\left. \frac{\partial^2 \ell(t)}{\partial \varepsilon^2} \right|_{\varepsilon=0}&=&
e^{i E_0 t}
2 \sum_{n=2}^{\infty} \frac{(-it)^n}{n!}
E_0^{n-2}
\left[
\sum_{k=0}^{n-2}
E_0^{-k} (n-1-k)
\sum_{\alpha=0}^{N-1}
 |V_{0\alpha}|^2 E_\alpha^k
\right].
\ea
where $V_{0\alpha}=\bra{\alpha}V \ket{0(\lambda)}$.
We can do the sum over $k$ first,
\ba
\left. \frac{\partial^2 \ell(t)}{\partial \varepsilon^2} \right|_{\varepsilon=0}&=&
e^{i E_0 t}
2 \sum_{n=2}^{\infty} \frac{(-it)^n}{n!}
E_0^{n}
\sum_{\alpha=0}^{N-1}
 |V_{0\alpha}|^2
\frac{n-1+\left(\frac{E_\alpha}{E_0}\right)^n-n\left(\frac{E_\alpha}{E_0}\right)
}{(E_\alpha-E_0)^2},
\ea
(notice that the term with $\alpha=0$ is finite), followed by the sum over $n$,
\ba
\left. \frac{\partial^2 \ell(t)}{\partial \varepsilon^2} \right|_{\varepsilon=0}&=&
2
\sum_{\alpha=0}^{N-1}  |V_{0\alpha}|^2
\frac{e^{-i (E_\alpha-E_0) t}-1+i t(E_\alpha -E_0)}
{(E_\alpha-E_0)^2}  \nonumber \\
&=&
-|V_{00}|^2 t^2 -
2
\sum_{\alpha=1}^{N-1}  |V_{0\alpha}|^2
\frac{1-e^{-i (E_\alpha-E_0) t}-i t(E_\alpha -E_0)}
{(E_\alpha-E_0)^2}
\label{mu2}
\ea

Now we need to put the results of Eqs. (\ref{mu0A}), (\ref{mu1}), and (\ref{mu2}) into Eq. (\ref{taylorA}),
\be
\ell(t)\simeq
1
 -i t V_{00} \varepsilon -
\left( |V_{00}|^2 t^2 + 2 \sum_{\alpha=1}^{N-1}  |V_{0\alpha}|^2
\frac{1-e^{-i (E_\alpha-E_0) t}-i t(E_\alpha -E_0)}
{(E_\alpha-E_0)^2} \right) \frac{\varepsilon^2}{2}.
\label{taylorB} \ee Using that $V_{00}$ is real and keeping the
term with lower order in $\varepsilon$, we obtain the expression
for the Loschmidt echo: \be L(t)=|\ell(t)|^2 \simeq 1 - 2
\varepsilon^2 \sum_{\alpha=1}^{N-1}  |V_{0\alpha}|^2 \frac{1-\cos
 (E_\alpha-E_0) t}{(E_\alpha-E_0)^2} \label{finalechoA} \ee

%----------------------------------------------
%----------------------------------------------

%----------------------------------------------
%----------------------------------------------

%----------------------------------------------
%----------------------------------------------

%----------------------------------------------
%--------------------------------------

%----------------------------------------------
%--------------------------------------

%----------------------------------------------
%----------------------------------------------

%----------------------------------------------
%----------------------------------------------

%----------------------------------------------
%----------------------------------------------

%----------------------------------------------
%----------------------------------------------

%----------------------------------------------
%----------------------------------------------

%%----------------------------------------------
%\begin{figure}
%\includegraphics[width=6in]{Lthermal}
%%Lthermal.m and Lthermal.fig in  F:\UW\PhaseTrans
%\caption{The Loschomidt echo in thermal state of three and four
%qubit systems shown as figures (a-b) for various temperature $T$
%shown in the figures. The critical points can be well indicated by
%the minima of the $L$ when the temperature is not too high, shown
%as the curves except the thick solid ones.} \label{Lthm}
%\end{figure}
%%----------------------------------------------


\begin{thebibliography}{}
\bibitem{QPTbook} S. Sachdev, {\it Quantum Phase Transitions} (Cambridge University Press, Cambridge,
                            2000);
                    M. Vojta, Rep. Prog. Phys. {\bf 66}, 2069 (2003);
                    P. Coleman and A. J. Schofield, Nature (London) {\bf 433}, 226 (2005).

\bibitem{entangedQPT} A. Osterloh, L. Amico, G. Falci, and R. Fazio, Nature (London)
                    {\bf 416}, 608 (2002); T. R. DeOliveira et al., Phys. Rev. Lett. {\bf 97}, 170401
                    (2006).

\bibitem{adiabatic} E. Farhi et al., Science, {\bf 292}, 472
                        (2001); J. I. Latorre and R. Or\'{u}s,
                        Phys. Rev. A {\bf 69}, 062302 (2004);
                        R. Sch\"{u}tzhold and G. Schaller, Phys. Rev. A {\bf 74}, 060304(R)
                        (2006).

\bibitem{Zanardi07} P. Zanardi, M. G. A. Paris, and L. C. Venuti, Phys. Rev. A {\bf 78}, 042105
                        (2008);
                    C. Invernizzi et al., Phys. Rev. A {\bf 78}, 042106
                    (2008).

\bibitem{RevNaturePhys08} S. Sachdev, Nature Phys. {\bf 4}, 173
                (2008); P. Gegenwart, Q. Si,  and  F. Steglich, {\it ibid.}, 186
                (2008); T. Giamarchi, C. R\"{u}egg, and O. Tchernyshyov, {\it ibid.}, 198 (2008).

\bibitem{sunPRL06} H. T. Quan, Z. Song, X. F. Liu, P. Zanardi, and C. P. Sun,
             Phys. Rev. Lett. {\bf 96}, 140604 (2006);
             Z.-G. Yuan, P. Zhang, and S.-S. Li, Phys. Rev. A {\bf 75}, 012102
             (2007).

\bibitem{zanardi:031123} P. Zanardi and N. Paunkovi$\acute{c}$, Phys. Rev. E {\bf 74}, 031123 (2006).

\bibitem{Paz} F. M. Cucchietti, S. Fernandez-Vidal, and J. P. Paz, Phys. Rev. A {\bf 75}, 032337 (2007).

\bibitem{Zhang08} J. Zhang et al., Phys. Rev. Lett. {\bf 100}, 100501
                    (2008).

\bibitem{IsingStat84} J. R. Kirkwood, Journal of Statistical Physics,
            {\bf 37}, 407 (1984).

\bibitem{compile} M. D. Bowdrey, J. A. Jones, E. Knill, and R.
                Laflamme, Phys. Rev. A {\bf 72}, 032315 (2005)

\bibitem{qchaos} J. Karthik, A. Sharma, and A. Lakshminarayan, Phys. Rev. A {\bf 75}, 022304
                        (2007).

\bibitem{pfeuty} K. Uzelac, R. Jullien, and P. Pfeuty,
Phys. Rev. B {\bf 22}, 436 (1980).

\bibitem{physRepIsing} M. A. Continentino, Phys. Rep. {\bf 239}, 179 (1994).

\bibitem{IsingQPT} P. Sen, Phys. Rev. E {\bf 63}, 016112 (2000).

\bibitem{Bose01Ising} D. Gunlycke, V. M. Kendon, V. Vedral, and S. Bose, Phys. Rev. A
                        {\bf 64}, 042302 (2001).

\bibitem{Zoller05} W. H. Zurek, U. Dorner, and P. Zoller, Phys. Rev. Lett. {\bf 95}, 105701
                                    (2005).

\bibitem{Werner05} P. Werner et al., Phys. Rev. Lett. {\bf 94},
                        047201 (2005).

\bibitem{Peng05} X. Peng, J. Du, and D. Suter, Phys. Rev. A {\bf 71}, 012307 (2005).

\bibitem{Mostame07} S. Mostame, G. Schaller, and R.
                    Sch\"{u}tzhold, Phys. Rev. A {\bf 76},
                    030304(R) (2007).

\bibitem{Gu07} S.-J. Gu et al., arXiv:0706.2495v2 [quant-ph].

\bibitem{Ovchinnikov} A. A. Ovchinnikov, D. V. Dmitriev, V. Ya. Krivnov, and
V. O. Cheranovskii, Phys. Rev. B {\bf 68}, 214406 (2003).

\bibitem{suzuki}
M. Suzuki, Prog. Theo. Phys. {\bf 56}, 1454 (1976).

\bibitem{IsingAF} E. M\"{u}ller-Hartmann and J. Zittartz, Z. Physik B {\bf 27}, 261 (1977);
K. Binder and D. P. Landau, Phys. Rev. B {\bf 21}, 1941 (1980).

\bibitem{supersymmetry1} J. Vidal, R. Mosseri, and J. Dukelsky, Phys. Rev. A {\bf 69}, 054101 (2004).

%\bibitem{supersymmetry2} E. Witten, Nucl. Phys. B {\bf 188}, 513 (1981). %Strange: in the paper, the Vol No. is 185

\bibitem{cqpt} S. L. Sondhi, S. M. Girvin,  J. P. Carini, and D.
                    Shahar, Rev. Mod. Phys. {\bf 69}, 315 (1997).

\bibitem{LEReview}  R. A. Jalabert and H. M. Pastawski, Phys. Rev. Lett. \textbf{86}, 246 (2001);
T. Gorin, T. Prosen, T. H. Seligman, and M. $\check{Z}$nidari$\check{c}$, Phys.
                Rep. {\bf 435}, 33 (2006).

\bibitem{Rossini} D. Rossini. T. Calarco, V. Giovannetti, S. Montangero, and R. Fazio,
                Phys. Rev. A {\bf 75}, 032333 (2007).
                    %e-print quant-ph/0611242.

\bibitem{You} W.-L. You, Y.-W. Li, and S.-J. Gu, Phys. Rev. E. {\bf 76}, 022101 (2007).

\bibitem{Geometric} L. C. Venuti and P. Zanardi, Phys. Rev. Lett. {\bf 99}, 095701
                            (2007).

\bibitem{Landau} L. D. Landau and E. M. Lifshitz, Quantum Mechanics
(Pergamon, London, 1958); C. Zener, Proc. R. Soc. London {\bf A137}, 696 (1932).


\bibitem{Damski05} B. Damski, Phys. Rev. Lett. {\bf 95}, 035701
                            (2005).

\bibitem{Poulin} D. Poulin and P. Wocjan, arXiv:0809.2705v1
                    [quant-ph].

\bibitem{Temperature} P. Zanardi, H. T. Quan, X. Wang, and C. P. Sun, Phys. Rev. A {\bf 75}, 032109 (2007).

\bibitem{Suter06} T. S. Mahesh and D. Suter, Phys. Rev. A {\bf 74}, 062312
                    (2006).

\bibitem{BellNMR}  A. M. Souza, et al., arXiv:0711.1156v2
                        [quant-ph].

%\bibitem{Yi07} L. C. Wang, X. L. Huang, and X. X. Yi, arXiv:0704.3463v2
%                [quant-ph].

%\bibitem{Miquel} C. Miquel, J. P. Paz, M. Saraceno, E. Knill,
%                    R. Laflamme, and C. Negrevergne, Nature, {\bf
%                    418}, 59(2002).

\bibitem{tce} M. A. Nielsen, E. Knill and R. Laflamme, Nature {\bf 396}, 52
                (1998).


\bibitem{grape}  J. Baugh et al., Physics in Canada, {\bf 63}, No. 4 (2007),
                    "Special issue on quantum information
                 and quantum computing",  also seeing arXiv:0710.1447v1 [quant-ph];
                    N. Khaneja et al., J. Mag. Res. {\bf 172},
                296 (2005);
                C.A. Ryan et al., Phys. Rev. A {\bf 78}, 012328 (2008).

\bibitem{refocus} N. Linden, \={E}. Kup\v{c}e and R. Freeman, Chem. Phys. Lett. {\bf
                311}, 321, (1999);
                 L. M. K. Vandersypen and I. L. Chuang,
                       Rev. Mod. Phys. {\bf 76},
                      1037 (2004).

\bibitem{effectivePure} D. G. Cory et al.,
                Physica D {\bf 120}, 82 (1998);
                X. Peng et al., arXiv: quant-ph/0202010;
                J. Zhang et al., Phys. Rev. A {\bf 76}, 012317
                (2007).


\bibitem{zhangpra04}  J. Zhang et al., Phys. Rev. A {\bf 70}, 062322
                (2004).


\bibitem{crot} E. Knill et al., Nature
                (London) {\bf 404}, 368 (2000);
                C. A. Ryan et al., Phys. Rev. Lett. {\bf 95}, 250502
                (2005).
%                J. Du et al., arXiv:0712.2694v1 [quant-ph].


\bibitem{CoryPure4} J. S. Hodges, P. Cappellaro, T. F. Havel, R. Martinez, and D. G.
                        Cory, Phys. Rev. A {\bf 75}, 042320 (2007).

\bibitem{weiCP} D.-X. Wei et al., Chinese Physics, {\bf 13}, 817
                        (2004).

\bibitem{swap} Z. L. Madi, R. Br$\ddot{u}$schweiler, and R. R. Ernst, J. Chem. Phys. {\bf
                    109}, 10603 (1998).


\bibitem{Wang08} X. Wang, Z. Sun, and Z. D. Wang, arXiv:0803.2940v2
                    [quant-ph].

\bibitem{Yang07} M.-F. Yang, Phys. Rev. B {\bf 76}, 180403(R)
                    (2007).

\bibitem{Quanpreparation} H. T. Quan and F. M. Cucchietti, arXiv:0806.4633v1 [quant-ph].

\end{thebibliography}
\end{document}